\def\year{2020}\relax
\documentclass[letterpaper]{article}
\usepackage{aaai20}
\usepackage{times}
\usepackage{helvet}
\usepackage{courier}
\usepackage{url}
\usepackage{graphicx}
\usepackage{gensymb}
\usepackage{amsmath}
\usepackage{siunitx}

\newcommand{\etal}{\emph{et.\ al.~}}

\title{Inferring Nighttime Satellite Imagery from Human Mobility}
\author{
Brian Dickinson,\textsuperscript{\rm 1}
Gourab Ghoshal,\textsuperscript{\rm 1}
Xerxes Dotiwalla,\textsuperscript{\rm 2}
Adam Sadilek,\textsuperscript{\rm 2}
Henry Kautz \textsuperscript{\rm 1}\\
\textsuperscript{\rm 1}University of Rochester\\
500 Joseph C. Wilson Blvd.\\
Rochester, New York 14627\\
\textsuperscript{\rm 2}Google Inc.\\
1600 Amphitheatre Parkway\\
Mountain View, California 94043\\
bdicken3@cs.rochester.edu, gghoshal@pas.rochester.edu, xerxes@google.com, 
sadilekadam@google.com, kautz@cs.rochester.edu
}

\begin{document}
\maketitle
\begin{abstract}
Nighttime lights satellite imagery has been used for decades as a uniform, global source of data for studying a wide range of socioeconomic factors. Recently, another more terrestrial source is producing data with similarly uniform global coverage: anonymous and aggregated smart phone location. This data, which measures the movement patterns of people and populations rather than the light they produce, could prove just as valuable in decades to come. In fact, since human mobility is far more directly related to the socioeconomic variables being predicted, it has an even greater potential. Additionally, since cell phone locations can be aggregated in real time while preserving individual user privacy, it will be possible to conduct studies that would previously have been impossible because they require data from the present. Of course, it will take quite some time to establish the new techniques necessary to apply human mobility data to problems traditionally studied with satellite imagery and to conceptualize and develop new real time applications. In this study we demonstrate that it is possible to accelerate this process by inferring artificial nighttime satellite imagery from human mobility data, while maintaining a strong differential privacy guarantee. We also show that these artificial maps can be used to infer socioeconomic variables, often with greater accuracy than using actual satellite imagery. Along the way, we find that the relationship between mobility and light emissions is both nonlinear and varies considerably around the globe. Finally, we show that models based on human mobility can significantly improve our understanding of society at a global scale.
\end{abstract}

\section{Introduction}
Nocturnal lighting is one of the most recognizable signs of human development on our planet. Perhaps the clearest sign of an area that has undergone human development is the appearance of electrical lighting. At night, these lights are detectable from orbit and have been collected for decades, first as a secondary application of military and weather satellites and more recently from new satellites with the primary mission of scientific inquiry \cite{hall2001history}. The launch of these new satellites was in response to a growing range of applications for this data because it had proven to be a reliable source of uniformly collected global data that doesn't suffer the limitations of census (e.g., limited coverage, bias, time lags) and is not subject to confounding factors like media influence and political corruption. Applications of this data are in a wide range of fields, from predicting socioeconomic factors such as poverty or GDP \cite{elvidge2012night,elvidge2009global} to estimating greenhouse gas emissions \cite{doll2000night}.

As time has passed, another data source is nearing the same level of global coverage with mostly uniform collection standards: smart phone location. We believe that many existing studies which use nighttime lights satellite data could benefit from an alternate data source based on human mobility rather than light output. This is especially true since intuitively many of the factors these studies seek to estimate have a  more direct relationship with human mobility than with electric light production. Additionally, human mobility from cell phones could be aggregated in real time and released very promptly, while annual nighttime satellite data releases tend to trail actual collection by at least 18 months. In this paper, we propose to use anonymous and aggregated flows from users opted-in to Google's Location History -- a global data source -- as an alternative data source for these studies. We do this by using human mobility to predict detected nighttime light worldwide. This allows for direct application of human mobility data to previously studied problems without individually adopting it to the diverse set of existing methodologies. In addition to creating high fidelity artificial maps, we show that in general our artificial satellite imagery is more highly correlated with GDP than real nighttime lights, demonstrating the benefits of using data based on human mobility rather than nocturnal light production.

\section{Background}
\subsection{Nightsat Data and Applications}

There are two major sources of nighttime satellite imagery. The older is the Operational Line-Scan System (OLS) system run by the Defense Meteorological Satellite Program (DMSP), first launched in 1961, declassified in 1972 and now being phased out as satellites fail without replacement \cite{hall2001history}. Its  successor, Visible Infrared Imaging Radiometer Suite (VIIRS), was launched in 2011. A comparison of the sensors, orbits, and other relevant features was performed by Elvidge \etal and shows that VIIRS data will likely be preferable for any studies that do not require time points earlier than 2011 \cite{elvidge2013viirs}. Our analysis will focus primarily on VIIRS data since it is the preferred system for post-2011 analysis and our mobility data begins in 2015. Our review of applications however will include work done with DMSP/OLS since researchers have had more time to develop novel applications with this dataset.

There have been four core datasets released based on DMSP/OLS data: daily/monthly, ``Stable Light'', ``Radiance-Calibrated'', and time-series. The daily, monthly, and time-series data is adjusted only to calibrate differences between satellites. For the stable lights dataset, additional steps are taken to remove ephemeral lights, e.g., fires and gas flares. Finally, the calibrated dataset attempts to correct for the sensor saturation which commonly appears in core urban areas. There are also several data products related to VIIRS: ``vcm'', ``vcm-nlt'', ``vcm-orm'', and ``vcm-orm-ntl''. In these product codes ``orm'' stands for outlier removed and ``ntl'' stands for nighttime lights. In this case, the ``orm'' designation indicates that a preprocessing step similar to the one used in the DMSP/OLS ``Stable Light'' data has been used to remove ephemeral lights. The nighttime lights designation is necessary because VIIRS unlike DMSP/OLS also provides daytime observations. 

Annual VIIRS satellite maps are released with significant delay, likely due to the increased processing required for aggregation and outlier correction \cite{elvidge2017viirs}. These maps are of higher quality than monthly releases and cannot be directly reproduced from a year's individual monthly maps. At the time of this writing (May 2019) the most recent set of annual maps available are for 2016. VIIRS data is also released monthly in the ``vcm'' and ``vcmsl'' configurations. The monthly ``vcm'' release, like its annual counterpart, does not correct for outliers or ephemerial lights. In the ``vcmsl'' product, some correction is made for stray light. This can provide more coverage towards the poles, but generally is of lower quality than the standard product. Monthly maps are released promptly about a month after collection.

Nighttime lights satellite imagery has been used for many applications in a wide range of areas, from tracking forest fires to estimating population density \cite{lo2001modeling,sutton2009paving,ebener2005wealth,chand2006monitoring,small2005spatial,doll2000night,elvidge2009global,elvidge2012night}. The most commonly used datasets have been daily/monthly data for short-term monitoring, and ``Stable Light'' for estimating urban extent and socioeconomic information \cite{huang2014application}. This makes sense as they are the finest timescale and cleanest regular-release products respectively. 
Many of these studies remain particularly relevant today, particularly those that focus on global poverty and greenhouse gas emissions. A decades old application of DMSP/OLS data which is perhaps more relevant today even than when it was first performed is mapping global greenhouse gas emissions, such as the study  by Doll \etal in 2000 \cite{doll2000night}. They focus on the strong correlations between nighttime light and GDP and between GDP and CO2 emissions. In essence, they converted measured light to predicted emissions by applying the linear correlation coefficients first between light and GDP and then between GDP and CO2 emissions. The resulting map of carbon emissions was quite similar to the CDIAC estimates for the same period. This is important because while still the gold standard for mapping carbon emissions, CDIAC maps are perpetually five years out of date due to the time it takes to collect and aggregate the data.

In 2008 Elvidge \etal developed a global poverty map based on DMSP/OLS nighttime lights data \cite{elvidge2009global}. They recognized that national level aggregation distorted poverty levels in some areas. To do this they combined DMSP/OLS light data with Landscan population estimates to identify areas with significantly lower light levels per person at a global scale with much finer spatial resolution. This is built on the idea of nighttime light production as a proxy for wealth, which has been supported by a number of other studies. Their actual metric divides the population of a 30 arc-second grid cell by its emitted light value. Next, the linear correlation coefficient for the sum of these values and the reported poverty index for each country was calculated allowing for a rough translation of their index to widely used poverty index measures. The result was a global poverty map with much finer spatial resolution and no bias from country boundaries.

In a followup work published in 2012, Elvidge \etal introduced a ``Night Light Development Index'' (NLDI) which uses the distribution of light among the population of an area to estimate the area’s level of development \cite{elvidge2012night}. In order to distill this distribution into a single metric they use the Gini coefficient, a common measure of inequality in the distribution of a resource based on the Lorenz curve. They demonstrate that their NLDI is negatively correlated with the Human Development Index (HDI) with the relatively strong coefficient of r2=0.71 . This suggests that NLDI might be a good surrogate measure for HDI at the sub-national level.

In this initial study we will focus on using human mobility data to predict VIIRS ``vcm-orm-ntl'' satellite imagery and show that this simulated satellite imagery can be used to predict GDP with similar or even superior accuracy. Additionally, time-sensitive studies could greatly benefit from significantly shortened waiting periods between data collection and release. We believe that these benefits extend to finer timescale studies which could benefit from our one-week artificial maps. In the future, we will show that many more applications could benefit from our simulated data.

We also believe that there are a great number of unexplored applications for human mobility data in general and our inferred satellite maps in particular. Most interesting to us is the potential for applications that make predictions about these socioeconomic variables in real time. Before our study, this would be impossible, however, we demonstrate that it is possible to infer accurate global  maps from only a week of smart phone location data, which could be aggregated and released much more quickly than VIIRS satellite data. Such applications could cover as wide a range of fields as the ones we have discussed here, including timely GDP estimates and tracking significant events.

\subsection{Google Location History Data}

The Google Mobility dataset is a heavily aggregated and anonymized summary of global trips mostly provided by location services on Android phones of users who have enabled location history. This dataset is perhaps the first to provide near-global mobility coverage using uniform definitions \cite{bassolas2019}.

This dataset uses S2 geometry which provides a hierarchical representation of the surface of a sphere by projecting it onto a bounding cube producing much less distortion than traditional map projections \cite{s2geometry}. This system provides a hierarchical set of cells. At the largest scale, $6$ level $0$ cells cover the entire surface of the earth. At each subsequent level the cells of the previous level are subdivided into $4$ additional cells, for example there are $24$ level $1$ cells. In total there are $31$ levels in the hierarchy with level $30$ cells each covering less than 1cm2.

The basic geographic units for this dataset are level $12$ and $13$ S2 cells. Level $12$ S2 cells cover areas ranging from \SI{3.04}{km^2} to \SI{6.38}{km^2} depending on latitude; level $13$ cells, similarly, cover areas of \SI{0.76}{km^2} to \SI{1.59}{km^2}. These resolutions are comparable to those provided by DMSP/OLS and VIIRS respectively. Each entry in the dataset can be formulated as a tuple of the form $a, b, t, n \pm e$ where $a$ and $b$ are the ids of the source and destination cells, $t$ is the time interval, and $n$ is the total number of trips made from cell $a$ to cell $b$ during the time interval $t$ with added Laplacian noise $e$. The included noise provides $(\epsilon, \delta)$-differential privacy where $\epsilon=0.66$ and $\delta=\num{2.1e-29}$. In other words, there is a $1-(2.1 \times 10^-29)$ probability that the inclusion of a user’s data in this dataset changes inferences about them by no more than $0.66\%$ relative to what could be inferred if their data was excluded from the dataset. This is a very strong differential privacy guarantee. Additionally, only tuples where $n > 100$ are included in the dataset in order to provide $k$-anonymity. This guarantees that any trip a user makes is included in the dataset only if at least 99 other people made a trip that is indistinguishable from theirs in our representation. This for example preserves the privacy of a single individual travelling to a remote location such as a private cabin during a particular week, which would still have been identifiable through our aggregation without the threshold providing k-anonymity. These heavy aggregations, privacy guarantees, and minimum trip thresholding maximize the individual privacy of Google's users while providing incredibly useful data for global population level analysis.

\section{Methods}
The most straightforward way to evaluate how well suited this mobility data is to many applications initially designed to work with DMSP/OLS and VIIRS satellite data is to predict satellite imagery as an intermediate step. The trade-off here is improved direct comparability and interoperability at the expense of some accuracy. We believe this trade-off worthwhile for our study as we seek to demonstrate the broad applicability of mobility data. However, in the future we hope to demonstrate the even greater potential of direct prediction. In order to predict light accurately we must take into account differences in the correlation between mobility and light across regions. We handle this in two ways: first using kriging, a well established technique in geospatial studies and secondly using a simpler technique using predefined regions \cite{oliver1990kriging,pebesma2006role}. In this second analysis, we defined our regions using the geographic subregions defined by the World Bank. In both analyses, we used non-linear regression with cross-validation to predict light values for the pixels in an artificial replication of VIIRS satellite imagery.

Our method of predicting a map of worldwide nighttime light emissions follows this basic outline, with step 3 varying somewhat depending on the chosen technique.
\begin{enumerate}
	\item Extract mobility metrics for all s2 cells with mobility data
	\item Extract actual mean light values for surface area represented by these cells
	\item Use cross-fold prediction with regression model to predict light for each cell
	\item Fill in predicted light values to all pixels represented by a cell (all others black)
\end{enumerate}

\subsection{Global and Segmented Regression Models}

\subsubsection{Extracting Relevant Light and Mobility Metrics}
The first step in training our regression model is to extract actual light values for each cell in our mobility data. This is done by taking the mean light intensity of the pixels in the satellite imagery that most directly correspond to each cell. Next, we extract a number of mobility features from the Google Location History dataset. For our analysis we use total out-flow, total self-flow, median trip distance, and total trip distance (for more information see Supplemental Material). Each of these metrics is statistically significant with $p < 0.01$ in our linear regression analysis. Due to the strong correlation between total out-flow and total in-flow we choose to use only out-flow in our analysis.These metrics are chosen to provide a concise picture of the unique flow characteristics of each cell.

\subsubsection{Initial Global Models}

Aside from the  computationally expensive task of matching cell mobility and light values for over 100 million S2 cells, which required several weeks of compute time, this is at its core a regression problem. In particular, the goal is to estimate the relationship between the mobility profile of a cell and its average light output. We first constructed a simple linear regression model to provide a baseline. Using $5$-fold cross-fold validation, performing linear regression on our 2016 annual dataset resulted in prediction a mean absolute error of $178.47$. Next, we performed the same analysis using a random forest regression model which surprisingly yielded a higher mean absolute error of $211.46$ \cite{breiman2001random,geurts2006extremely}. In terms of our particular regression problem, this indicates that for any given cell our radiance prediction was off by around \SI{200}{nanowatts \per cm^2 \per sr} on average (where the unit ``sr'' is square radians). For context, the mean radiance of cells with measurable light cell is \SI{2.55}{nanowatts \per cm^2 \per sr} with a standard deviation of \SI{25.47}{nanowatts \per cm^2 \per  sr}. The distribution of cell radiance values is heavily skewed towards zero with only $11.6\%$ of cells having any detectable radiance. With error rates this high, landmasses and major cities would be identifiable, but the inferred maps would be of little use for any real applications. Additionally, upon further evaluation of our models, it became apparent that the error values of both the linear and random forest models varied considerably between folds.

This was surprising because preliminary analysis using latitude longitude grid patches had shown much lower error rates. These patches were much smaller than the global analysis, but were still quite substantial covering approximately $5$ million square kilometers each. The significantly higher error rates on global models indicate that the relationship between mobility patterns and light emissions varies somewhat more than we expected in different parts of the world, but is much more reliably predictable on a smaller scale.

Because we found that in different parts of the world the relationship between mobility and light production may differ considerably, our models must account for such regional variations. We explore two potential solutions to this problem. The first solution is to use kriging to create a meta-model of spatial variations in this relationship. Second, we propose a simpler solution utilizing predefined regions which we find performs similarly well with much lower complexity.

\subsubsection{Regression-Kriging}
Regression-Kriging is used to predict the spatial distribution of a dependant variable using both spatial information and one or more environmental variables that are correlated with the dependant variable \cite{pebesma2006role}. In this application light is the dependant variable to be predicted and our mobility metrics are the related environmental variables. Spatial information is easily inferred from the centroid of each cell. One complication of this strategy is the need for greater care in partitioning training and testing data. Without great care, models are able to take advantage of spatial-autocorrelation in the dependant variable to significantly improve performance in random cross-fold validation resulting in significant under reporting of error rates \cite{roberts2017cross}. To combat this, we make two significant changes to our cross-fold validation procedure. First we segment all of our data into over $50,000$ $1\degree$ longitude by $1\degree$ longitude blocks, of which about $7,000$ actually have some mobility information. Learning models for each of these blocks separately significantly limits the proximity of cells in the training and testing sets. Our second change additionally excludes blocks adjacent to predicted block from the training set. These changes should be sufficient to eliminate the inadvertent bleed-through of information from the spatial-autocorrelation of the dependant variable and produce much more accurate error estimates without overfitting, however, it is difficult to be absolutely certain all outside influences have been properly addressed. Due to the increased computational complexity of this technique, we were forced to model only a $7\degree$ by $7\degree$ area for each predicted block. This should not significantly impact results as it as a relatively large area that we expect to be quite similar to the predicted area. Using regression-kriging for prediction, our error rate was reduced to a mean absolute error of $7.43$, a dramatic improvement over either of our naive global models.

There are two significant downsides to this technique: the difficulty of designing countermeasures to completely eliminate overfitting and the sheer number of parameters required to fit $7,000$ separate regression models. In the hopes of mitigating these concerns, we also construct simpler models for a much smaller number of predefined regions.

\subsubsection{Regional Models}
Our simpler solution instead defines separate models for different fixed geographic areas. We define fourteen regions based on World Bank geographic subregions: Northern America, Latin America and the Caribbean, Eastern Europe, Southern Asia, Southeast Asia, Southern Europe, Western Europe, Western Asia, Northern Europe, Eastern Asia, Sub-Saharan Africa, Northern Africa, Australia and New Zealand, and Central Asia. Because of their small size and proximity, we chose to merge Southeast Asia, Polynesia, Melanesia, and Micronesia into a single region. The first step of our segmentation process is to assign cells to their regions which are defined as sets of countries.

We begin by assigning all level $12$ and $13$ S2 cells for which we have mobility data into the countries they overlap. This is done using country boundary definitions from GADM version 2.8. In the simplest and most common case, an S2 cell belongs to exactly one country and is assigned the region of that country. Cells that span multiple countries that are all part of the same region are similarly assigned to that region. These two cases account for over $99\%$ of cells. Cells that span international borders between regions or appear outside of national borders (most commonly along coastlines where GADM polygons are insufficiently precise) are assigned regions using the following technique. Cells that have already been assigned regions are used to compute convex hulls for each of the regions. Any unassigned cell that is contained in exactly one of these hulls is assigned to the corresponding region. Finally, all remaining cells are assigned to the region of their five nearest neighbors. With all of our S2 cells assigned to a region we can finally begin training our models.

After completing this regional segmentation of our cells, we repeated the linear and random forest model cross-fold validation experiment for each World Bank Sub-Region individually. In this case block cross-fold prediction is not preferred to random cross-fold prediction, because these regression models do not make use of spatial information. The results for these regional models were quite promising. The linear and random forest models for the North American region, which is relatively well covered with mobility data, had mean absolute errors of $23.70$ and $18.63$ respectively. These error rates compare much more favorably with the standard deviation of the data. Unsurprisingly, the models for the less well covered Southeastern Asia had errors of $108.15$ and $108.00$. These errors are, however, much lower than for full worldwide models and indicate relatively accurate predictions that may produce useful inferred maps. Because our random forest regressors had considerably lower total error and variations in error between folds, we decided to use random forest models in all further analysis.

\subsubsection{Reconstructing Artificial Maps}
Regardless of which technique is used, our regression models produce predicted light values for each cell with measured mobility. In order to reconstruct a proper map, the predicted light value for each cell where mobility is available is applied to every pixel covered by that cell. All cells with no mobility information are assumed to have no light value. This process generates maps strikingly similar to actual satellite data. In fact, despite the seemingly significant differences in regression error between regression-kriging and our simpler regional models we see very similar quality maps. For example, using regression-kriging with our ``All-Weeks'' dataset our inferred map has a mean absolute error of $0.06$ and a mean squared error of $31.22$. The corresponding errors for our regional models are $0.06$ and $31.32$. Furthermore if we use these metrics as a measure of the difference between our two inferred maps, we find that they are even more similar to each other than they are to the observed imagery (MAE: $0.02$; MSE $0.59$). For this reason, and due to our concerns about overfitting, we have chosen to perform the majority of our analyses using the simpler regional models, though we also include some kriging results for comparison.

While a high fidelity map is desirable and indicates the effectiveness of our models, it is important to remember that our ultimate goal is not to perfectly reproduce satellite imagery. Instead, our objective is to actually improve on the original imagery by providing information about human mobility which we will show is a better indicator of GDP. Therefore, some of the ``error'' in our predictions may actually be desirable indicating levels of mobility that are unusually high or low for the amount of light emitted. In order to demonstrate this we repeat one of the most basic analyses performed on nighttime satellite imagery, GDP to light correlations. In this analysis we show that the correlations with our predicted light values are similar to and sometimes stronger than those with actual light values.

\section{Results and Conclusions}

We performed the above analysis to reproduce VIIRS-ORM-NTL imagery from 2015 and 2016 from several Google Location History datasets. These datasets are enumerated below and provide a picture of the applicability of fine-grained temporal timescales. We show that in areas where there is sufficient coverage, these artificial maps show a stronger correlation between light emissions and GDP. In cases where there is very strong coverage, we find that the ``All-Weeks'' dataset produces the most highly correlated maps likely due to its finer spatial scale (Level $13$ cells average \SI{1.27}{km^2} compared with the \SI{5.07}{km^2} resolution of the level $12$ cells in our ``Annual'' dataset). As examples of these correlations in densely covered areas, see correlations in the United States in  and Tables \ref{tab:state_correlations} and \ref{tab:covered_country_correlations} (Scatter Plots for USA available in Supplemental Materials). In areas with sparser coverage, we find that the ``Annual'' dataset produces more highly correlated maps, likely due to the increased data available when applying the hundred trip minimum thresholding on total number annually and on larger source and destination cells. Only in very sparsely covered areas, such as Sub-Saharan Africa, were the correlations stronger in the original imagery (See Supplementay Material).
In addition to the ``Annual'' and ``All-Weeks'' datasets we also included several subsets of the ``All-Weeks'' data with finer timescales. These datasets perform almost as well as the ``All-Weeks'' data and show potential for more fine-grained temporal analysis. Most strikingly, even maps produced from a single week of mobility data are nearly as accurate as those using an entire year of data. This can be seen in Tables \ref{tab:state_correlations} and \ref{tab:covered_country_correlations}.

\begin{itemize}
	\item Annual - Level $12$ S2 cells with Laplacian noise and $k$-anonymity applied to annual total
	\item All-Weeks - Level $13$ S2 cells with Laplacian noise and $k$-anonymity applied to each week
	\item Spring - Subset of ``All-Weeks'' including only weeks occurring during astronomical spring (between the vernal equinox and summer solstice)
	\item May - Subset of ``All-Weeks'' including only weeks in the month of May
	\item Week-20 - Subset of ``All-Weeks'' including only the twentieth week of 2016 which occurred in May
\end{itemize}

\begin{figure*}[!htbp]
  \includegraphics[width=\textwidth]{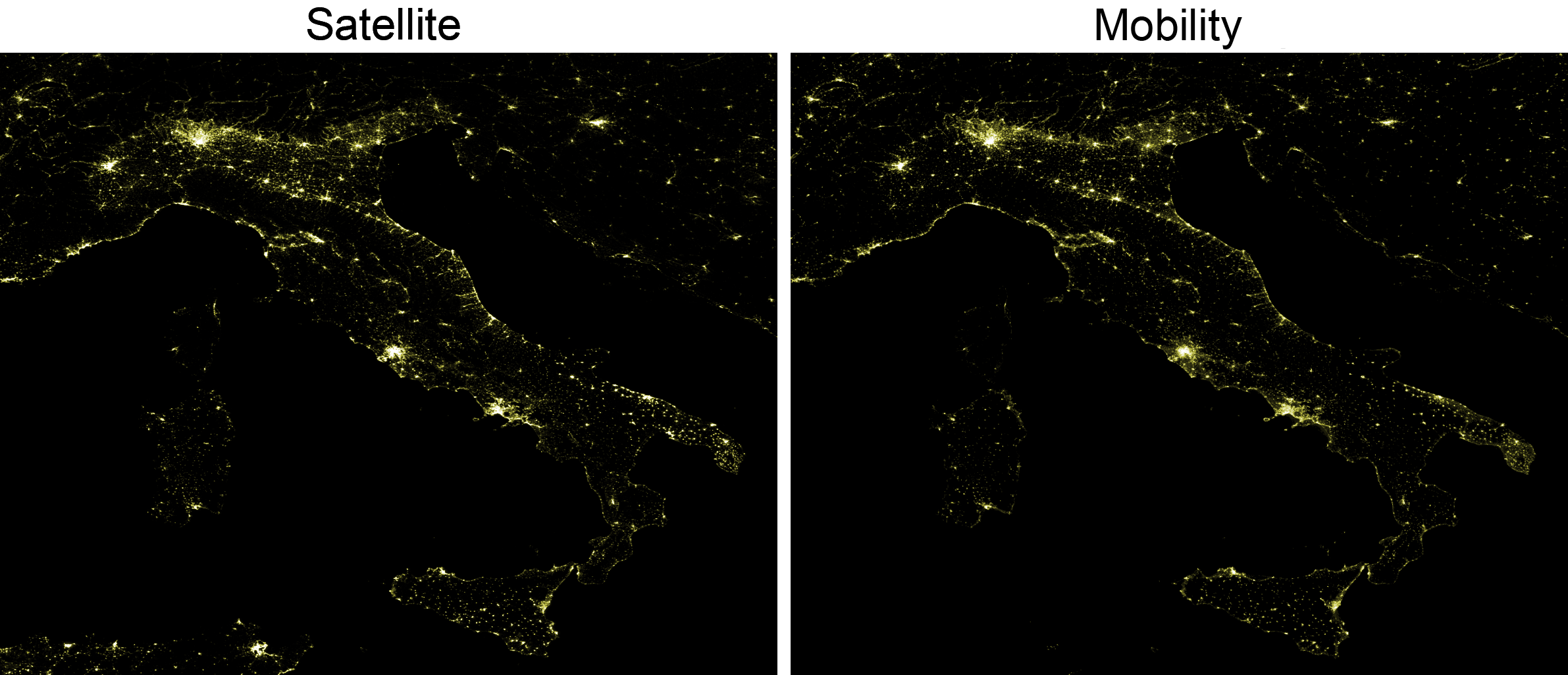}
  \caption{Inferred nighttime lights satellite imagery. The mobility map was constructed without any satellite data -- it was inferred from human mobility data based on the anonymous and aggregated flows of users opted-in to Google's Location History (our ``All-Weeks'' dataset). Here we show visual comparison between it and a light map from Visible Infrared Imaging Radiometer Suite (VIIRS). There are, of course, some minor differences between the maps; most notably  the northern coast of Tunisia and Algeria shows less activity in mobility than in light. As we have seen, the light map can be inferred from mobility data at a comparable level of resolution and accuracy, but can be done in real-time, while preserving strong user privacy, rather than with the 18 month delay for annual satellite products or the noise of monthly releases. This added timeliness and scale opens new research and application avenues.}
  \label{fig:italy_maps}
\end{figure*}

\begin{figure*}[!htbp]
    \centering
    \includegraphics[width=0.85\textwidth]{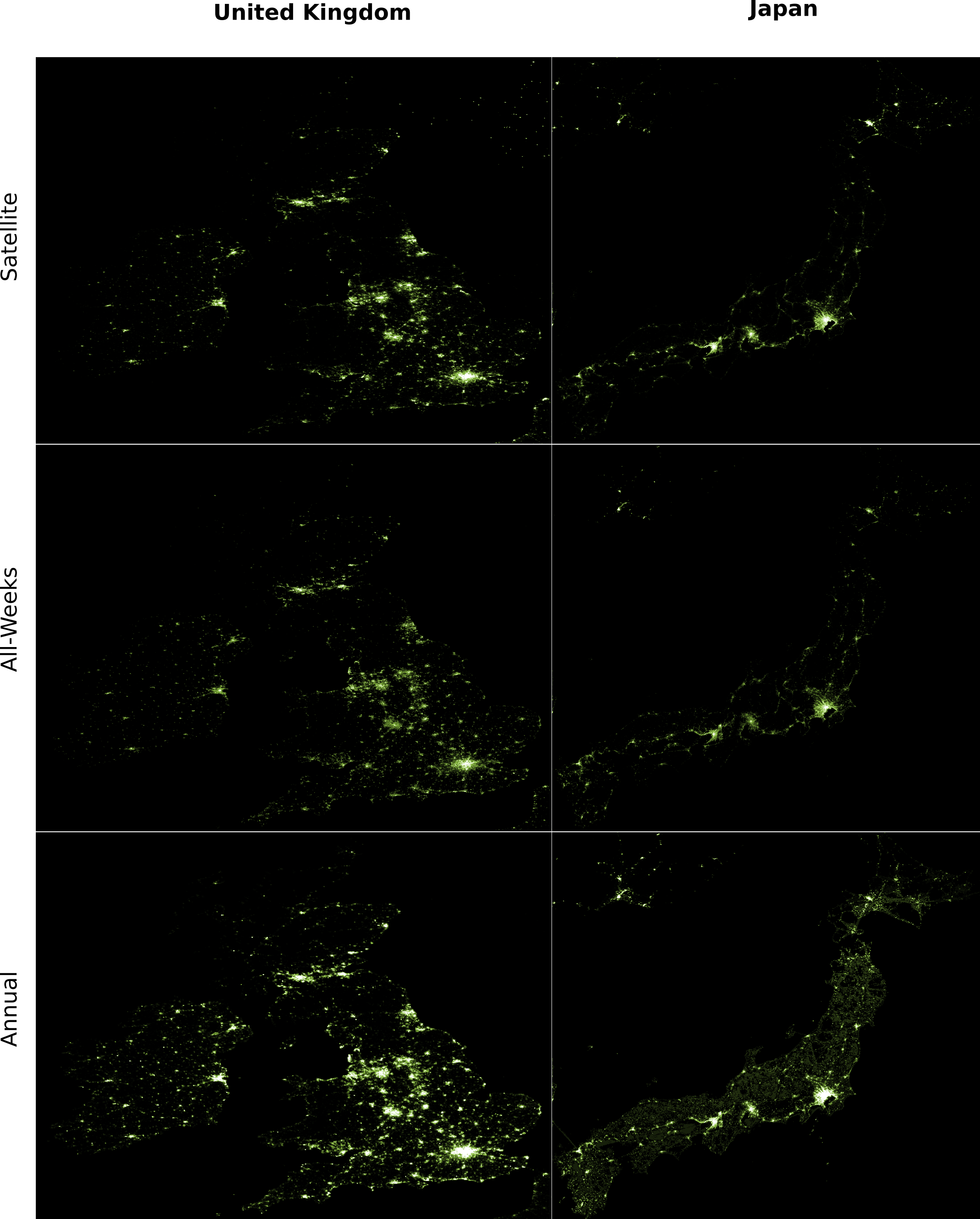}
    \caption{Actual and inferred nighttime lights imagery for the United Kingdom and Japan in 2016. The top row ``Satellite'' is taken directly from the VIIRS outlier removed nighttime lights product. The middle row ``All-Weeks'' is in an artificially generated map based on weekly aggregations of mobility data with about a $1.25 km^2$ spatial resolution for each cell. The bottom row ``Annual'' is another artificial map predicted from an annual aggregation of mobility data with a spatial resolution of about $5 km^2$ for each cell. As a result the annual map is somewhat coarser, however it also has greater global coverage since the $100$ trip threshold is applied for the entire year rather than for every individual week. Both mobility datasets are provided by Google location history. The ``All-Weeks'' predictions in particular are visually remarkably similar to the real data. While less like the original maps, the predicted ``Annual'' imagery provides additional potentially useful highlighting that might be missed in either of the other maps.}
    \label{fig:uk_japan_maps}
\end{figure*}

\begin{figure}
    \centering
    \includegraphics[width=0.85\columnwidth]{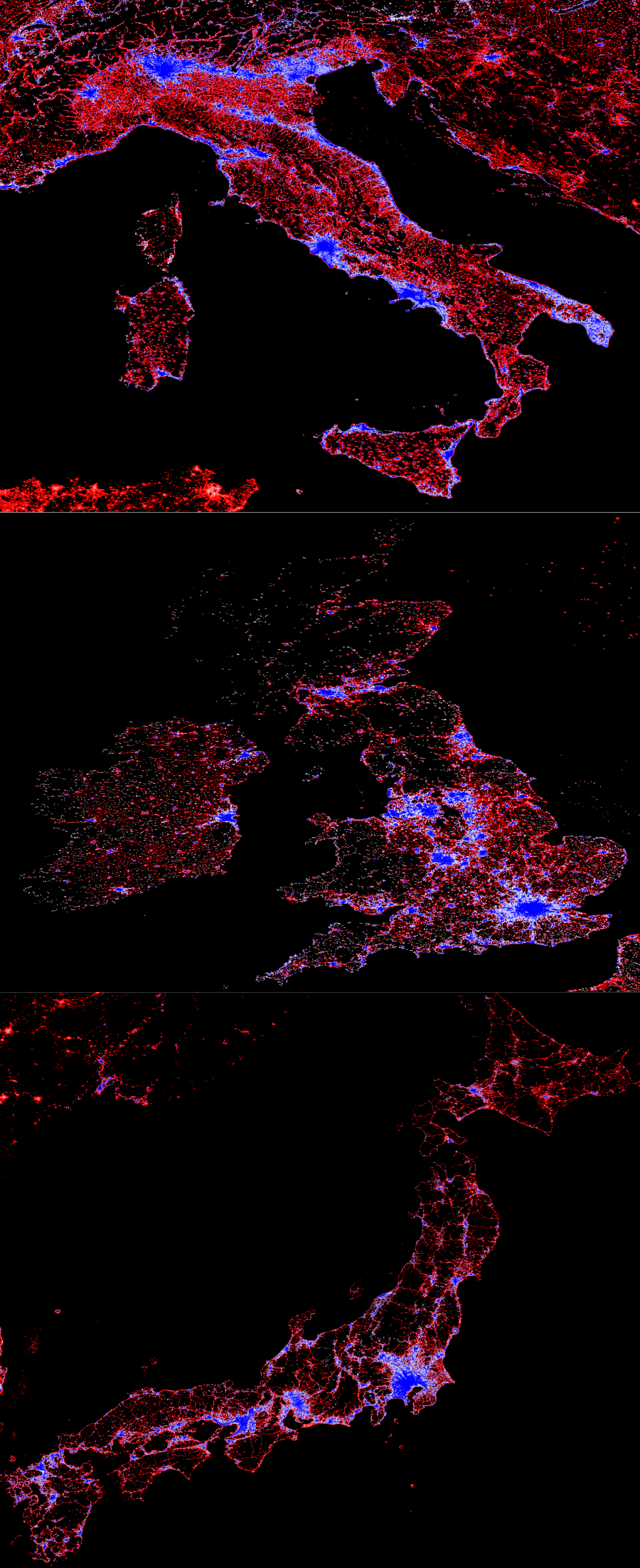}
    \caption{These plots highlight the minor, but systematic differences between actual and mobility inferred nighttime satellite imagery (2016 VIIRS and ``All-Weeks'' predicted). In these plots, areas with no data are black and areas with equal measured and inferred light are white. The more mobility outweighs light in an area, the more blue the area will appear. Conversely the more light outweighs mobility, the more red the area will appear. In these plots one immediately notices that mobility data emphasizes urban and coastal areas, and under represents rural areas. This is most likely due to the $k$-anonymity thresholding that is applied to mobility data and not satellite data.}
    \label{fig:diff_maps}
\end{figure}

Overall the simulated satellite imagery is strikingly similar to actual imagery, with a mean absolute error of $0.06$ in our ``All-Weeks'' imagery. The level of similarity is visually demonstrated in Figures \ref{fig:italy_maps} and \ref{fig:uk_japan_maps}. Additionally, since human mobility is often a better proxy for socioeconomic variables than light emission, the places where these maps differ often correspond to areas where predictions from actual light emissions would over or underestimate the desired variables or where our dataset is more privacy preserving (Figure \ref{fig:diff_maps}). This shows great opportunities for improving on the results of many existing studies simply by using similarly predicted maps in place of actual satellite imagery. This would, however, of course only be a first step toward applying mobility data to applications previously analyzed with satellite data. While the creation of light maps allows for trivial translation of existing work, providing a direct method for comparison to existing studies and for replicating such studies with more timely data, it does not exhaust the potential of the mobility datasets. Our purpose here is to demonstrate concisely and simply the potential benefits of using global human mobility patterns alongside satellite imagery to open up new opportunities for study.

\begin{table*}
	\centering
	\caption{Spearman correlation between GDP and Total Light for states in each region of the United States. Statistical significance is indicated by the number of asterisks following the correlation. Three indicate that $p < 0.001$, two that $p < 0.01$, and one that $p < 0.05$. If no asterisks follow the correlation, then it is not statistically significant. Note that in all regions the artificial maps perform at least as well as the real data. Of particular note are the Southern and particularly the Western United States where the correlation is substantially higher with all artificial maps than with the real map. Also of interest is that across all regions only a single week of data is needed to reach peak performance. This is likely due to the completeness of mobility coverage for the United States, but shows the power of these methods for short-timescale analysis.}
	\begin{tabular}{|c|c|c|c|c|c|}
	\hline
		\textbf{Map} & \textbf{All} & \textbf{Northeast} & \textbf{Midwest} & \textbf{South} & \textbf{West} \\ \hline
		2016 Real & $0.8546^{***}$ & $0.9667^{***}$ & \textbf{$0.9720^{***}$} & $0.8701^{***}$ & $0.7802^{**}$ \\ \hline
		2016 Annual & $0.9375^{***}$ & \textbf{$0.9833^{***}$} & $0.9580^{***}$ & $0.9461^{***}$ & $0.9505^{***}$ \\ \hline
		2016 All-Weeks & $0.9598^{***}$ & \textbf{$0.9833^{***}$} & \textbf{$0.9720^{***}$} & \textbf{$0.9632^{***}$} & \textbf{$0.9670^{***}$} \\ \hline
		2016 Spring & $0.9584^{***}$ & \textbf{$0.9833$} & $0.9580$ & \textbf{$0.9632^{***}$} & \textbf{$0.9670^{***}$} \\ \hline
		2016 May & $0.9591$ & \textbf{$0.9833^{***}$} & \textbf{$0.9720^{***}$} & \textbf{$0.9632^{***}$} & \textbf{$0.9670^{***}$} \\ \hline
		2016 Week-20 & \textbf{$0.9639^{***}$} & \textbf{$0.9833^{***}$} & \textbf{$0.9720^{***}$} & \textbf{$0.9632^{***}$} & \textbf{$0.9670^{***}$} \\ \hline
		2016 Kriging & $0.9086^{***}$ & $0.9833^{***}$ & $0.9580^{***}$ & $0.9510^{***}$ & $0.8352^{***}$ \\ \hline 
	\end{tabular}
	\label{tab:state_correlations}
\end{table*}

\begin{table*}
	\centering
	\caption{Spearman correlation between GDP and Total Light for countries in well covered regions. Statistical significance is indicated by the number of asterisks following the correlation. Three indicate that $p < 0.001$, two that $p < 0.01$, and one that $p < 0.05$. If no asterisks follow the correlation, then it is not statistically significant. As this table shows, the correlation with artificial maps is quite similar to that of the actual light maps for many well covered areas. The only major exception is the ``All'' column where the real data outperforms our artificial maps. It is important to note that the correlations in this column are global and that most of the performance loss in the artificial maps comes from either very sparsely covered areas such as Central Asia and Sub-Saharan Africa or areas where mobility data could not be collected or released (such as  China). Otherwise, the only true outlier in performance is in Northern Europe, where all our artificial maps performed quite well, but the actual light map does not. Again, note that the performance provided by maps generated with only a single week of mobility data do not substantially underperform any of the other maps.}
	\begin{tabular}{|c|c|c|c|c|c|c|c|c|c|}
	\hline
		\textbf{Map} & \textbf{All} & \textbf{W. Europe} & 
		\textbf{E. Europe} & \textbf{N. Europe} & \textbf{S. Europe} &
		\textbf{L. America} & \textbf{S. Asia} \\ \hline
		2016 Real & \textbf{$0.9384^{***}$} & $0.8571^{*}$ & \textbf{$0.9429^{**}$} & $0.6167^{*}$ & \textbf{$0.9860^{***}$} & $0.9610^{***}$ & $0.9286^{***}$ \\ \hline
		2016 Annual & $0.8738^{***}$ & $0.8571^{*}$ & \textbf{$0.9429^{**}$} & \textbf{$0.9000^{***}$} & $0.9720^{***}$ & $0.9610^{***}$ & \textbf{$1.0000^{***}$} \\ \hline
		2016 All-Weeks & $0.8420^{***}$ & \textbf{$0.8929^{**}$} & \textbf{$0.9429^{**}$} & $0.8833^{**}$ & $0.9790^{***}$ & \textbf{$0.9651^{***}$} & $0.9762^{***}$ \\ \hline
		2016 Spring & $0.8362^{***}$ & \textbf{$0.8929^{**}$} & $0.8857^{*}$ & $0.8833^{**}$ & $0.9510^{***}$ & $0.9624^{***}$ & $0.9762^{***}$ \\ \hline
		2016 May & $0.8320^{***}$ & \textbf{$0.8929^{**}$} & $0.8857^{*}$ & $0.8833^{**}$ & $0.9510^{***}$ & $0.9590^{***}$ & $0.9762^{***}$ \\ \hline
		2016 Week-20 & $0.8300^{***}$ & \textbf{$0.8929^{**}$} & \textbf{$0.9429^{**}$} & $0.8833^{**}$ & $0.9510^{***}$ & $0.9576^{***}$ & $0.9762^{***}$ \\ \hline
		2016 Kriging & $0.8334^{***}$ & $0.7857^{*}$ & $0.9643^{***}$ & $0.8333^{**}$ & $0.9720^{***}$ & $0.9398^{***}$ & $1.0000^{***}$ \\ \hline
	\end{tabular}
	\label{tab:covered_country_correlations}
\end{table*}

\section{Discussion and Future Work}

In this study, we have demonstrated it is possible to reconstruct nighttime lights satellite maps from as little as a single week of human mobility data. This opens up the possibility of applying these maps for fine-grained time-series analyses based on weekly changes in measured mobility. Furthermore, the predicted light values in these maps are generally more highly correlated with socioeconomic factors than are the actual measured light emissions. Additionally, since human mobility as measured by smart phone location can be aggregated quickly, these artificial maps could be made available more promptly than those provided by the VIIRS satellite system. Therefore, we believe that human mobility based artificial maps show great promise in many applications that have previously used nighttime lights satellite imagery.

As part of this analysis we also demonstrated that the relationship between human mobility and nocturnal light emissions is both nonlinear and varies considerably around the globe. The differences across regions are made clear by the improvement in the performance when modeling each region independently rather than constructing a single global model. Similarly, the nonlinear relationship is demonstrated by the improved performance of our random forest regressor over classic linear regression. While it is beyond the scope of this study to analyze the ways in which light and mobility are related in each region, the simple existence of such differences could be an important factor in future research.

Since the quality of mobility based artificial maps will have an enormous impact on their utility, we plan to further refine our methods. This would include more thorough hyper-parameter tuning as well as the evaluation of a number of state of the art regression techniques including XGBoost, LightGBM, and fully connected artificial neural networks. Another avenue for expanding on this work is the application of our artificial maps to a number of other applications which have traditionally used nighttime satellite imagery. In particular applications constructing global poverty maps, development indices, and greenhouse gas emissions might be improved by the use of human mobility measures rather than nocturnal light production \cite{doll2000night,elvidge2012night,elvidge2009global}. This would further demonstrate the wide applicability of our work.

Finally, the end goal of our research is to demonstrate that models based on aggregate human mobility can improve our understanding of society at a global scale. While our success in light prediction immediately shows that this is the case, we are working on creating even stronger models for inferring socioeconomic metrics from mobility data.

\section{Acknowledgements}
We thank Avi Bar, Curt Black, Susan Cadrecha, Stephanie Cason, Charina Chou, Katherine Chou, Iz Conroy, Liz Davidoff, Jeff Dean, Damien Desfontaines, Paul Estham, Bryant Gipson, Jason Freidenfelds, Vivien Hoang, Sarah Holland, Michael Howell, Ali Lange, Onur Kucuktunc, Allie Lieber, Bhaskar Mehta, Caitlin Niedermeyer, Genevieve Park,  Ludovic Peran, Flavia Sekles, Aaron Stein, Chandu Thota and Ashley Zlatinov for their insights and guidance.

\bibliographystyle{aaai}
\bibliography{AISI-DickinsonB.191}
\end{document}